\documentclass[letterpaper,english,reprint,aps,notitlepage,superscriptaddress,twocolumn,pra]{revtex4-2}

\usepackage{lineno}

\usepackage[utf8]{inputenc}
\usepackage{amsmath}
\usepackage{graphicx}
\usepackage{xcolor}
\usepackage[separate-uncertainty=true]{siunitx}
\usepackage{braket}
\usepackage{hyperref}
\usepackage{url}
\usepackage[dvipsnames]{xcolor}
\colorlet{mylinkcolor}{NavyBlue}
\colorlet{mycitecolor}{NavyBlue}
\colorlet{myurlcolor}{NavyBlue}
\hypersetup{
	linkcolor  = mylinkcolor,
	citecolor  = mycitecolor,
	urlcolor   = myurlcolor,
	colorlinks = true,
	breaklinks = true
}

\newcommand{\nion}{n_{\text{ions}}}
\newcommand{\nm}[1]{\SI{#1}{\nano\meter}}
\newcommand{\alfasiunits}{\mega\hertz\square\meter\per\square\volt}
\newcommand{\probeline}{red}
\newcommand{\rydbergline}{magenta}
\renewcommand{\paragraph}[1]{}

\newcommand{\cent}{\affiliation{Centre for Quantum Optical Technologies, Centre of New Technologies, University of Warsaw, S. Banacha 2c, 02-097 Warsaw, Poland}}
\newcommand{\fuw}{\affiliation{Faculty of Physics, University of Warsaw, L. Pasteura 5, 02-093 Warsaw, Poland}}

\begin{document}

\title{Measuring Interaction-Induced Energy Shifts of Rydberg Atoms in Hot Vapor}

\author{Tomasz Prokop}
\email{t.prokop@cent.uw.edu.pl}
\cent
\fuw
\author{Bartosz Kasza}
\cent
\fuw
\author{Wojciech Wasilewski}
\cent
\fuw
\author{Michał Parniak}
\cent
\fuw

\begin{abstract} 
We demonstrate a method to measure energy shifts of the top level in a four-level ladder setup induced by atom interactions in thermal vapors. It utilizes the observation of two transmission minima corresponding to a split electromagnetically induced absorption (EIA) effect. We apply this method to measure mean Rydberg atom interactions in a hot vapor. We believe this approach could provide a valuable tool for accurately modeling mean-field Rydberg atom interactions, as well as sensing the occurrence of strong interactions.
\end{abstract}

\maketitle

\section{Introduction}
\paragraph{Introduction about Rydberg atoms}
Rydberg atoms -- alkali atoms excited to high principal number states -- are sensitive to electric fields, which makes them a promising platform for radio-frequency (RF) sensing and metrology~\cite{Fan_2015, Simons2021, Yuan_2023, Zhang_2024, Schlossberger2024}.
Several readout schemes have been demonstrated, including continuous wideband photon conversion~\cite{Borwka_2023}, superheterodyne detection~\cite{Jing_2020, Borwka_2025}, and homodyne readout~\cite{Kumar_2017_2}.
Rydberg-based receivers have also been applied in imaging radar~\cite{Watterson2025} and proposed for space applications~\cite{whitepaper}.

Characterizing the baseline performance of such receivers requires knowing whether the atomic response is affected by interactions between Rydberg atoms.

\paragraph{Interactions - bistability and applications}
Strong interactions between Rydberg atoms lead to significant shifts in the Rydberg level energy, which are foundational to intrinsic optical bistability in the atomic vapors. This phenomenon has been leveraged for enhanced metrology applications~\cite{Wang2026,Xue2026,Ding2022,Wu2024,Wang2023,Wade2018,Weichman2025}, time crystalline phase formation~\cite{Xue2026,XWu2024, Wu2026, Jiao2025, Wadenpfuhl2023,BLiu2025,BLiu2024,Gambetta2019, Arumugam2026, He2020}, and critical phenomena research~\cite{Carr2013,Zhang2024,Ding2024,Ding2020} among other applications~\cite{Liu2024,Liu2026}.

\paragraph{Prior Investigations}
Numerous studies have investigated intrinsic optical bistability~\cite{deMelo2016,ibali2016,Marcuzzi2014,Zhang2025,Ma2024,Lee2012} and the strong interactions that drive it~\cite{Weller2019, Weller2016, Wang2025}. For a precise study of the observed effects, a method is required for measuring the Rydberg level energy shift originating from strong atomic interactions in the hot vapors.

\paragraph{Method}
We present a method for measuring atomic level energy shifts using balancing of split EIA minima in a four-level atomic ladder setup.
These EIA minima are especially sensitive to coupling laser detuning; when the laser is on-resonance, the minima are ``balanced" - they have the same value of transmission. Conversely, when the laser is off-resonance, the probe transmission values in these minima diverge.
When the energy of the top (Rydberg) level of the ladder is shifted as a result of some interaction, the balance - equal transmission value in these EIA minima - is lost. In the method presented in this manuscript, this level energy shift is compensated with detuning the coupling laser in order to balance the EIA minima. After the balancing, the compensating detuning is equal to the top level energy shift.

\paragraph{Apply the method}
We apply this split EIA minima balancing method to hot vapors of rubidium Rydberg atoms in order to assess the energy shift of the Rydberg level due to self-interaction for different Rydberg atoms number densities.

\paragraph{Foreshadow results}
Previous studies considered van der Waals interactions and DC-Stark detuning originating from ionized Rydberg atoms as the origin of strong interactions. 
We compare our results with theoretical estimations for both Stark detuning from ions as well as van der Waals interactions. 
We find that there is a correspondence to the prediction for Stark detuning from ions, while van der Waals interactions exhibit a different character. This conclusion is in agreement with considerations existing in the literature~\cite{Weller2016}.

\section{Theoretical model}

\begin{figure*}
\centering
\includegraphics[width=\linewidth]{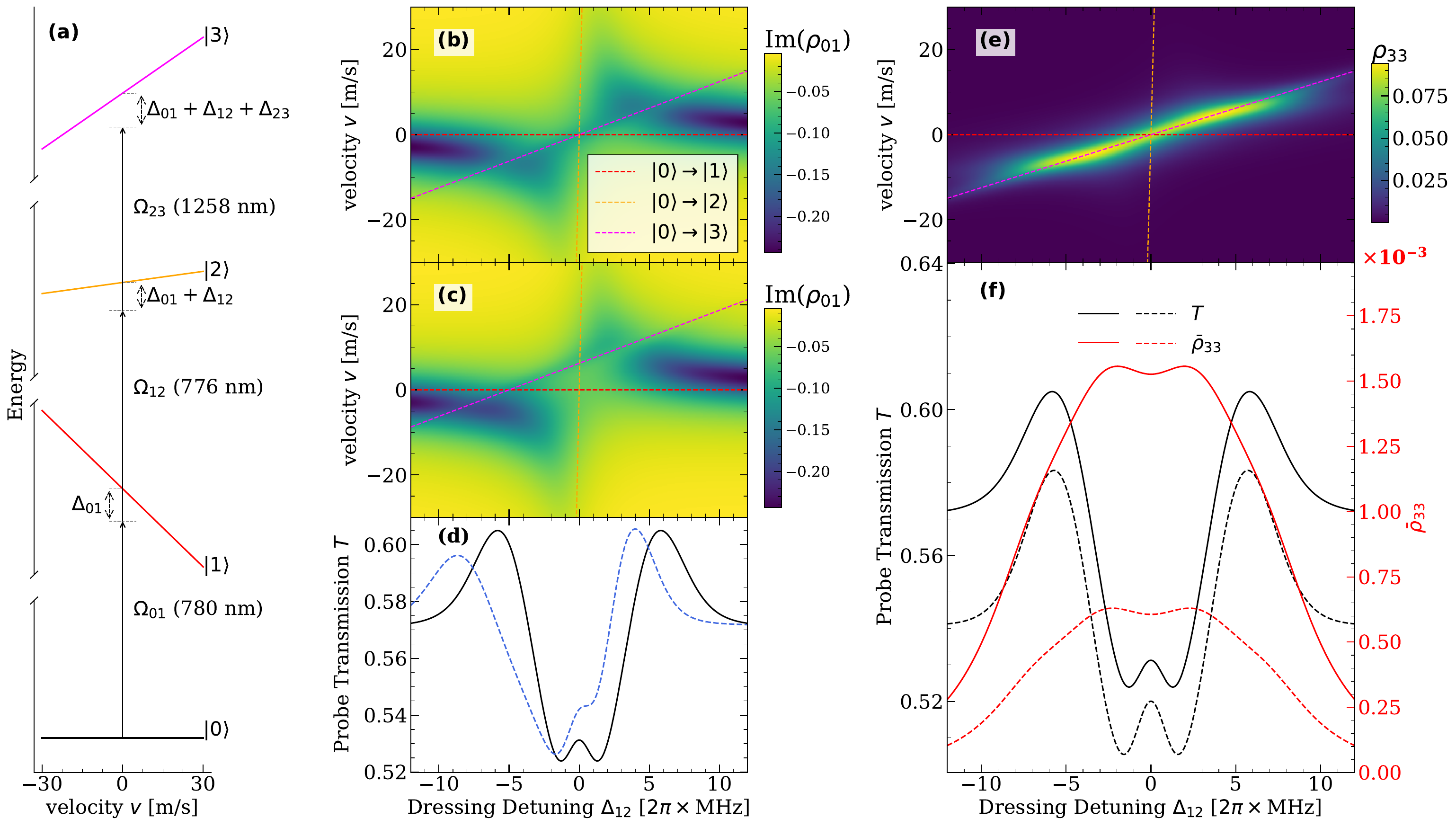}

\caption{Balancing the split EIA in a four-level ladder.
\textbf{(a)} Energy levels diagram.
The gray dashed lines indicate the value of detunings for stationary atoms.
The solid lines depict the Doppler-shifted energies, and their energy over velocity slopes are given by the cumulative sum of wave vectors.
\textbf{(b)} and \textbf{(c)} Colormaps of absorptive coherence Im($\rho_{01}(\Delta_{12},v)$) for resonant coupling laser ($\Delta_{23} = 0$) \textbf{(b)} and shifted Rydberg level energy $E_3$ i.e. detuned coupling laser ($\Delta_{23} = 2\pi \times \SI{5}{\mega\hertz}$). \textbf{(c)}
Dashed, colored lines follow $\ket{0}\rightarrow\ket{i}$ multi-photon resonances (c.f. Eq.~\eqref{eq:resonances}).
The \probeline\;line for the probe transition ($\ket{0}\to\ket{1}$) is horizontal due to scanning the dressing laser detuning $\Delta_{12}$.
A decrease of $E_3$ equates an increase of $\Delta_{23}$ and moves the three-photon resonance line $\ket{0}\to\ket{3}$ (\rydbergline) towards top-left.
\textbf{(d)} Intensity transmission of the probe laser for both the resonant ($\Delta_{23} = 0$, black solid line) and detuned ($\Delta_{23} = 2\pi \times \SI{5}{\mega\hertz}$, blue dashed line) coupling laser examples from above (\textbf{(b)}, \textbf{(c)}).
The transmission is calculated from Doppler-averaged $\bar\rho_{01}$ as explained in Eq.~\eqref{eq:transmission}.
\textbf{(e)} Colormap of Rydberg state population $\rho_{33}(\Delta_{12},v)$. The Rydberg state is populated along the \rydbergline\; resonance line as the result of the multi-photon resonance $\ket{0}\to \ket{3}$.
\textbf{(f)} Comparison of Doppler-averaged Rydberg state population $\bar\rho_{33}(\Delta_{12})$ (red, right y-axis) and probe transmission (black) for two probe laser Rabi frequencies $\Omega_{01}$ (solid lines: $2\pi\times \SI{5}{\mega\hertz}$, dashed lines: $2\pi\times \SI{3}{\mega\hertz}$), both for resonant coupling laser.
Increasing $\Omega_{01}$ preserves the profile of both probe transmission and population. 
The Rydberg state population value (red) is almost constant around the EIA minima, here around \SI{1.50e-3}{}.
}
\label{fig:theoretical_framework}
\end{figure*}

\paragraph{Levels}
Our method is applicable to a four-level ladder atomic system, such as depicted in Fig.~\ref{fig:theoretical_framework}(a). 
In this manuscript the relevant $^{87}$Rb atom's levels are: $\ket{0}=\ket{5^2S_{1/2}\;F=2, m_F=2}$, $\ket{1}=\ket{5^2P_{3/2}\;F=3, m_F=3}$, $\ket{2}=\ket{5^2D_{5/2}\;F=4, m_F=4}$ and $\ket{3}=\ket{55^2P_{3/2}\;m_j=3/2}$ - the Rydberg level. 
Optical transitions are driven by the probe laser \nm{780}, $\ket{0} \to \ket{1}$, 
dressing \nm{776}, $\ket{1} \to \ket{2}$ 
and coupling \nm{1258}, $\ket{2} \to \ket{3}$.
The Rabi frequencies and laser detunings will be denoted as $\Omega_{01}$, $\Omega_{12}, \Omega_{23}$ and $\Delta_{01}, \Delta_{12}, \Delta_{23}$ respectively. 

These levels are the most relevant because the transitions between the states with maximal angular momentum projection have the biggest magnitudes of transition dipole moments.

\paragraph{Hamiltonian and resonances}
To model the system, we utilize a semi-classical Hamiltonian in the interaction picture $H_I$ following~\cite{Krokosz2025, Kasza2025}, 
which includes Rabi frequencies of respective transitions as off-diagonal elements, $\langle m|H_I|n \rangle=\hbar\Omega_{nm}/2, n\neq m$, and the values of cumulative detunings of multi-photon transitions on the diagonal $\langle n|H_I|n \rangle$.
For a non-zero velocity of the atom $v$, these cumulative detunings include Doppler shifts of the frequencies of each of the lasers:

\begin{align}
    \langle n|H_I|n \rangle &= \hbar\sum_{i=1}^{n} (\Delta_{i-1,i} + k_{i-1,i} v), \label{eq:resonance} \\
    \Delta_{i-1,i} &= \omega_{i-1,i} - \frac{E_i - E_{i-1}}{\hbar}, \label{eq:detunings}
\end{align}
where $E_i$ is the $\ket{i}$ level energy, $\omega_{i-1,i}$ is the frequency of the laser coupling $\ket{i-1}$ with $\ket{i}$, $k_{01}\equiv - k_{780}$, $k_{12}\equiv k_{776}$, and $k_{23}\equiv k_{1258}$ are the wavenumbers of the probing, dressing, and coupling laser beams respectively, and $\hbar$ is the reduced Planck constant.

A change in the Rydberg level energy $E_3$, due to the inter-atomic interactions, is equivalent to changing the coupling laser detuning $\Delta_{23}$ by the same value in the opposite direction, as seen from Eq.~\eqref{eq:detunings}. In the simulations, we only adjust $\Delta_{23}$.

A multi-photon transition resonance occurs on $\ket0\to\ket{n}$ transition when $\langle n|H_I|n \rangle = \langle 0|H_I|0 \rangle$ i.e. when detunings match. 
This yields three resonance conditions, given by the following equations:
\begin{align}
\ket{0}\to\ket{1}:~ &-k_{780}v - \Delta_{01} = 0,\notag \\ 
\ket{0}\to\ket{2}:~ &(k_{776} - k_{780})v - \Delta_{01} - \Delta_{12} = 0,\label{eq:resonances} \\ 
\ket{0}\to\ket{3}:~ &(k_{1258} + k_{776} - k_{780})v - \Delta_{01} - \Delta_{12} - \Delta_{23} = 0 \notag.
\end{align}

The above equations describe slanted lines in Fig.~\ref{fig:theoretical_framework}(a) in the energy level diagram. Note that $\ket{0} \to \ket{2}$ resonance has close-to-zero Doppler slant, because the \nm{780} and \nm{776} beams counter-propagate.

We find that these resonance lines form a basis for interpreting the simulation of the atomic state and observations of probe transmission, which will be presented later on.

\paragraph{Density matrix calc}
We describe the equilibrium state of the atomic system with a velocity-dependent density matrix $\rho(v)$, which we find using common tools of open quantum systems~\cite{Krokosz2025,Kasza2025}. We account for atomic decays and transit time broadening~\cite{Fan_2015} using the respective jump operators. From these and the Hamiltonian of the system $H_I(v)$, which is velocity-dependent, we construct a Lindblad operator $\mathcal{L}(v)$. For a specific velocity $v$, we solve the Lindblad equation for the steady-state, $\mathcal{L}(v)\rho(v) = 0$. We then take an average over the Maxwell-Boltzmann velocity distribution $\text{MB}(v)$ of the thermal vapor to obtain the average density matrix of the atoms $\bar\rho = \int dv \rho(v) \text{MB}(v)$.

\paragraph{Absorption}
The transmission of the probe laser is governed by the paraxial wave equation with polarization of the atomic medium on the respective transition, which is given by the atomic coherence $\rho_{01}$ between the levels $\ket{0}$ and $\ket{1}$. Assuming the linear response regime $\rho_{01}\propto \Omega_{01}$, which holds for a weak probe, we can integrate the wave equation. The result for the probe intensity transmission $T$ through the atomic medium is the Lambert-Beer law:
\begin{equation}
T = \exp \left( 2L k_{780} \frac{P}{\epsilon_0 E} \right), \quad \quad P= n d_{01} \text{Im}(\bar\rho_{01}),
\label{eq:transmission}
\end{equation}
where $P$ is the absorptive part of the atomic polarization, $E$ is the incident probe electric field amplitude, $n$ is the density of the atoms, $d_{01}$ is the $\ket{0}\to\ket{1}$ transition dipole moment, $L$ is the length of atomic medium and $\epsilon_0$ is vacuum permittivity.

\paragraph{Anti-crossings and EIT}
Figs.~\ref{fig:theoretical_framework}b and \ref{fig:theoretical_framework}c present colormaps of the absorptive component of optical coherence Im($\rho_{01}(\Delta_{12},v)$) as a function of detuning of the dressing laser $\Delta_{12}$ and atomic velocity $v$.

The aforementioned resonances, given by Eq.~\eqref{eq:resonances}, are presented on the colormaps as dashed lines.
Note that the maps have the velocity $v$ now on the y-axis; thus, the resonance lines are drawn at different angles compared to the energy levels diagram. Moreover, the energy y-axis from the diagram is transformed into $\Delta_{12}$ on the x-axis. This preserves the character of the $\ket{0}\to\ket{2}$ and $\ket{0}\to\ket{3}$ resonance lines, but according to Eq.~\eqref{eq:resonances}, the \probeline\; resonance line will now be flat at a constant velocity value - zero for a resonant probe.

The maps of Im($\rho_{01}(\Delta_{12},v)$) are complex, but to a large degree they follow anti-crossings of the resonance lines.
For our Rabi frequencies, the widest anti-crossing is due to the dressing laser $\Omega_{12}$ and takes place between the resonance lines $\ket{0}\to\ket{2}$ and $\ket{0}\to\ket{1}$ resulting in the probe absorption veering from the $\ket{0}\to\ket{1}$ resonance line towards the $\ket{0}\to\ket{2}$ resonance line, thus forming two distinct absorptive branches.
The distance between these branches corresponds to the splitting in the transmission spectrum, resulting in the electromagnetically induced transparency (EIT) effect.

\paragraph{EIA and its splitting} 
The Rydberg coupling laser $\Omega_{23}$ adds the $\ket{0}\to\ket{3}$ resonance line. Two more, minor anti-crossings occur where the $\ket{0}\to\ket{3}$ resonance line intersects with the two absorptive branches forming EIT. Absorption cumulates in the area between the absorptive branches, along the $\ket{0}\to\ket{3}$ resonance line, between the minor anti-crossings. This produces electromagnetically induced absorption (EIA).

The probe transmission scans over dressing detuning $T(\Delta_{12})$, shown in Fig.~\ref{fig:theoretical_framework}(d), are obtained by velocity-averaging of Im($\rho_{01}(\Delta_{12},v)$) (over $v$-axis in Fig.~\ref{fig:theoretical_framework}(b) and \ref{fig:theoretical_framework}c) and calculating transmission scans $T(\Delta_{12})$ from the averaged Im($\bar\rho_{01}(\Delta_{12})$), according to Eq.~\eqref{eq:transmission}.
These scans have two minima around $\Delta_{12} = 0$, which correspond to the split EIA - the focal points of our method.

For the observation of these minima, it is important to have a strong enough dressing Rabi frequency $\Omega_{12}$ compared to $\Omega_{23}$ and the decay rate of the $\ket{1}$ level. If too little dressing power is used, then a single EIA minimum is seen, without the desired splitting.

\paragraph{Loss of ``balance''}
A decrease of the Rydberg level energy $E_3$ results in an increase of the detuning of the coupling laser $\Delta_{23}$, according to Eq.~\eqref{eq:detunings}. 
This moves the $\ket{0}\to\ket{3}$ resonance line (\rydbergline) towards negative dressing laser detunings $\Delta_{12}$, as described by Eq.~\eqref{eq:resonances}, and visible when comparing Figs.~\ref{fig:theoretical_framework}b and \ref{fig:theoretical_framework}c.
The resonance lines are no longer symmetric with respect to changing the sign of $\Delta_{12}$ as showcased in the colormaps.
The minor anti-crossing at positive $\Delta_{12}$ is now further from the $\ket{0}\to\ket{1}$ resonance line, which weakens the corresponding EIA minimum, leading to an asymmetric transmission profile seen in Fig.~\ref{fig:theoretical_framework}(d) (blue).

The colormaps and plots illustrate how the movement of resonance lines in the $(\Delta_{12}, v)$ plane dictates the spectral behavior of the system. A shift of the energy of the top (Rydberg) level also shifts the detuning of the coupling laser $\Delta_{23}$ by the same amount. The predictable response of the EIA minima, when varying $\Delta_{23}$, illustrates the fundamental mechanism for sensing of the level energy shifts.

\paragraph{The takeaway - EIA minima}
The technique we present is based on the following observation: the balance of the EIA minima is highly sensitive to the Rydberg level energy, so its position can be tracked by re-tuning the coupling laser to restore that balance.

\paragraph{Rydberg state population}
A Rydberg state population colormap $\rho_{33}(\Delta_{12},v)$ is presented in Fig.~\ref{fig:theoretical_framework}(e).
It illustrates that the Rydberg state is populated along the $\ket{0} \to \ket{3}$ resonance line (\rydbergline).

In Fig.~\ref{fig:theoretical_framework}(f) the Doppler-averaged probe transmission $T(\Delta_{12})$ (black) and Rydberg state population $\bar\rho_{33}(\Delta_{12})$ (red, right y-axis) are visible for two different probe Rabi frequencies $\Omega_{01}$ (solid and dashed lines).

The spectral shapes of the scans of $\bar\rho_{33}(\Delta_{12})$ and $T(\Delta_{12})$, shown in Fig.~\ref{fig:theoretical_framework}(f), are mostly scaled by varying $\Omega_{01}$ with minor shape changes. 
Therefore, adjusting the probe Rabi frequency $\Omega_{01}$ is a good way to control the Rydberg population $\bar \rho_{33}$ at the EIA minima.

\paragraph{The takeaway - Rydberg population}
Notably, the Rydberg population $\bar \rho_{33}$ has a near-constant value around the EIA minima. 
This is important when we expect the Rydberg level energy to be linked to Rydberg population, 
because the EIA minima balance can be restored in a condition of known and relatively constant Rydberg population, thereby enabling robust measurements. 

\begin{figure}
\centering
\includegraphics[width=\linewidth]{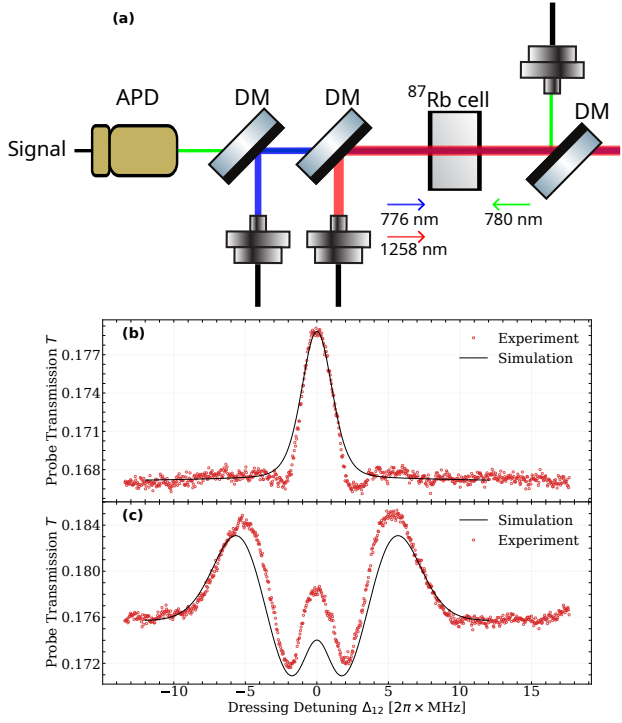}
\caption{\textbf{(a)} The experimental setup. APD - avalanche photodiode, DM - dichroic mirror. Probe transmission is measured via APD. The probe laser beam (\nm{780}) is aligned counter-propagating to the dressing (\nm{776}) and coupling (\nm{1258}) laser beams, and the beams are combined on DMs.
\textbf{(b)} and \textbf{(c)} Comparison of calibrated simulation with experimental data for calibration: \textbf{(b)} three-level resonance - without the \nm{1258} laser (EIT scan), \textbf{(c)} four-level resonance (split EIA scan).}
\label{fig:calibrations_setup}
\end{figure}

\section{Experimental setup}
\paragraph{The experimental setup}
The experimental setup scheme is presented in Fig.~\ref{fig:calibrations_setup}(a). The $^{87}\text{Rb}$ medium in the cell is \SI{7.5}{\milli\meter} long and has its temperature stabilized by a heater.
The probe laser beam is counter-propagating with respect to the dressing and coupling laser beams, in order to reduce the Doppler shift. Probe transmission is measured on the avalanche photodiode APD (Thorlabs APD410A/M).

\paragraph{Lasers}
All three laser beams are focused in the middle of the cell to a waist of $\SI{405 \pm 10}{\micro\meter}$. The beam powers incident on the cell are as follows: \nm{780}: $\SI{0.490 \pm 0.022}{\micro\watt}$ for simulation calibration measurements, and for measuring the interaction-induced energy shifts - from $\SI{0.490 \pm 0.022}{\micro\watt}$ to $\SI{3.91 \pm 0.18}{\micro\watt}$, \nm{776}: $\SI{0.92 \pm 0.028}{\milli\watt}$, \nm{1258}: $\SI{356 \pm 18}{\milli\watt}$, measured with Thorlabs photodiode power sensors: S130C for \nm{780} and \nm{776} and S132C for \nm{1258}.

All three lasers are frequency-locked. The probe (\nm{780}) and dressing (\nm{776}) lasers are frequency-modulated with acousto-optic modulators (AOM) and supplied by optical fibers to the system. Additionally, the power of these two lasers is locked using these AOMs. The dressing laser detuning scan $\Delta_{12}$ is performed through its AOM. All beams are circularly polarized to address the strongest transitions: $\sigma^+$ for the probe and dressing transitions and $\sigma^-$ for the coupling transition.

\paragraph{Atom density}

Alongside the relevant levels described in the previous section, more levels contribute to probe absorption, 
particularly in the absence of the dressing laser or on the sides of the transmission scans, away from the EIT/EIA features.

For the ground state, there are angular momentum projection levels with $m_F =\{-2,-1,0,1\}$ of the $5S_{1/2}\;F=2$ level.
Each of these ground state levels $\ket{g}$ has corresponding atomic density $n_{\ket{g}}$ and is coupled to at most 3 hyperfine states $\ket{e}$ of the $5P_{3/2}$ level,
each of different $F=1,2,3$ and an angular momentum projection of $m_F + 1$ due to the $\sigma^+$ polarization of the probe light.
Different $F$ manifolds of $5P_{3/2}$ also have different energies, yielding different probe detunings for subsequent transitions.
Each $\ket{g} \rightarrow \ket{e}$ transition has a different dipole moment $d_{ge}$, yielding a Rabi frequency $\Omega_{ge}$ for the common probe field.
For each transition, an equilibrium velocity-averaged density matrix $\bar\rho_{ge}$ can be calculated, taking into account these two levels alone.
Their contributions to the polarization add up: $P = \sum\limits_{g,e} n_{\ket{g}} d_{ge} \;\text{Im}(\bar\rho_{ge})$
and the expression for transmission Eq.~\eqref{eq:transmission} holds.

In the simulation, we describe the atomic ensemble using two different atomic densities: 
a density of the $\ket{0}$ level, $n_{\ket{0}}$ for $m_F=2$ 
and a density $n_{\ket{g}} = n_{\neq\ket{0}}$ for each of the other four states $m_F =\{-2,-1,0,1\}$.
In particular, near the $F=2 \rightarrow F'=3$ resonance, away from the EIT/EIA features,
$P/\epsilon_0 E \approx {(n_{\ket{0}} + 1.79 n_{\neq\ket{0}})/(\SI{5.33e20}{\per\meter\cubed})}$.

\paragraph{Calibrated parameters}
We calibrate the simulation parameters through iterative comparison of simulation and experimental data for the two cases presented in Fig.~\ref{fig:calibrations_setup}(b) - EIT scan and Fig.~\ref{fig:calibrations_setup}(c) - split EIA scan.
The Rabi frequencies are calculated for experimentally measured Gaussian beam peak intensity values and then scaled down by a factor common to all Rabi frequencies. This way, the spatial inhomogeneity of the Gaussian beams is accounted for. Scaled values are used in the simulation.

The common scaling factor of $0.67$ for Rabi frequencies and transit time broadening of $2\pi\times\SI{1.45}{\mega\hertz}$ are set to match the widths of peaks and troughs in the experimental data.
The atomic densities $n_{\ket{0}}$ and $n_{\neq\ket{0}}$ are set to match the transmission values at the sides of the transmission scans and the amplitude of the peaks and troughs.
The corresponding atomic densities chosen are: $n_{\ket{0}} = \SI{1.40e16}{\per\meter\cubed}$, $n_{\neq\ket{0}} = \SI{1.99e16}{\per\meter\cubed}$ for the EIT scan and $n_{\ket{0}} = \SI{8.46e15}{\per\meter\cubed}$, $n_{\neq\ket{0}} = \SI{2.22e16}{\per\meter\cubed}$ for the split EIA. The latter value of $n_{\ket{0}}$ enables estimating the Rydberg atom density.
Due to optical pumping to the $5S_{1/2}\;F=1$ level, the total atomic density in $F=2$ changes between measurements, and so does the $n_{\ket{0}}$.

\section{Results: Measuring interaction-induced energy shifts}

\begin{figure*}
\centering
\includegraphics[width=\linewidth]{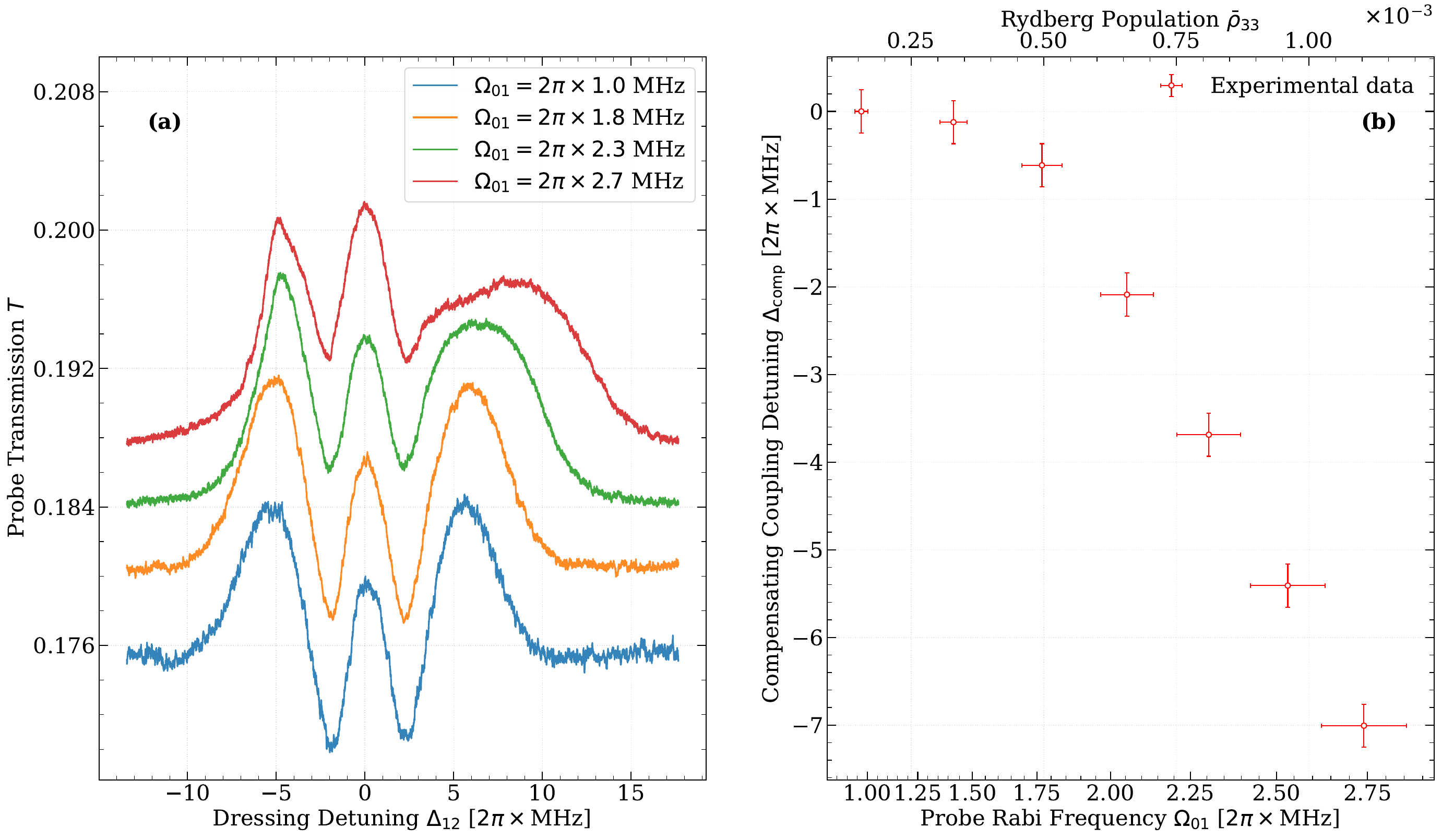}
\caption{
\textbf{(a)} Examples of observed probe transmission scans in $\Delta_{12}$ for different probe Rabi frequencies $\Omega_{01}$ after recovering the minima balance by tuning the coupling laser frequency by $\Delta_{\text{comp}}$, as indicated in the legend. 
\textbf{(b)} Coupling laser compensation tuning $\Delta_{\text{comp}}$ for set probe Rabi frequencies $\Omega_{01}$ (bottom x-axis).
For each $\Omega_{01}$, the Rydberg population $\bar\rho_{33}$ around the EIA minima (top x-axis) was calculated with the simulation.}
\label{fig:results}
\end{figure*}

\paragraph{Measurement results}
When varying the probe Rabi frequency $\Omega_{01}$ in the experiment, we observe disruption of minima balance similar to Fig.~\ref{fig:theoretical_framework}(d). 
We attribute the observed effect to an atomic interaction-induced Rydberg level energy shift.

To measure the Rydberg level energy shift, we restore the balance of the split EIA minima.
We reestablish the balance by shifting the coupling laser frequency by $\Delta_\text{comp}$.
Compensation is verified by the transmission scans shown in Fig.~\ref{fig:results}(a). 
This way, for each probe Rabi frequency $\Omega_{01}$, we find the compensation detuning $\Delta_{\text{comp}}$ as presented in Fig.~\ref{fig:results}(b). 
The amount of compensation $\Delta_{\text{comp}}$ is equal to the Rydberg level energy shift due to atomic interactions.

\paragraph{Population from simulation}
From the calibrated simulations, we obtain the value of $\bar \rho_{33}$ around the split EIA minima (as exemplified in Fig.~\ref{fig:theoretical_framework}(f)) for each probe Rabi frequency $\Omega_{01}$.
Combined with experimental data, we recover the energy shift $\Delta_\text{comp}$ as a function of the Rydberg state population $\bar\rho_{33}$ depicted in Fig.~\ref{fig:results}(b) with the help of the top x-axis.

\section{Discussion}

\paragraph{Overview}
We compare our results for the Rydberg level energy shift, $\Delta_{\text{comp}}$, with theoretical predictions.
We consider two possible causes: ionization or van der Waals interactions.

\begin{figure*}
\centering
\includegraphics[width=\linewidth]{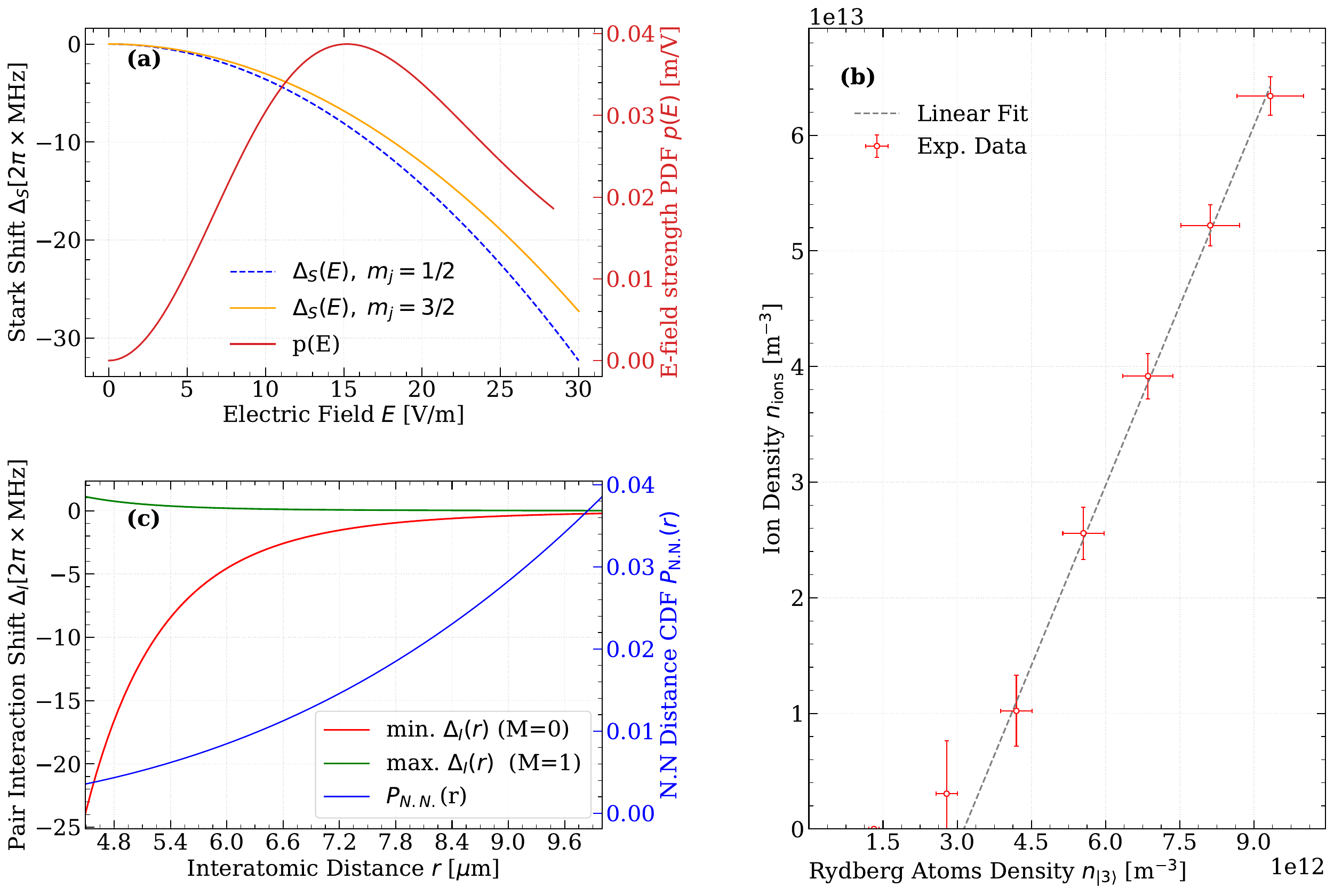}
\caption{
\textbf{(a)}
Stark shift $\Delta_S(E)$ curves for the $\ket{55P_{3/2}\;m_j=3/2}$ state (orange, solid) and additionally for the $\ket{55P_{3/2}\;m_j=1/2}$ state (blue, dashed).
Electric field distribution (red, right y-axis) for $\nion = \SI{6.34e13}{\per\meter\cubed}$. For such an ion density, the Stark shift value for the median electric field is $2\pi\times \SI{7}{\mega\hertz}$ - the highest one in the results in Fig.~\ref{fig:results}(b).
\textbf{(b)}
The ion number density $\nion$ causing the $\ket{55P_{3/2}\;m_j=3/2}$ level energy shift $\Delta_{\text{comp}}$ 
against Rydberg number density $n_{\ket{3}} = n_{\ket{0}}\bar\rho_{33}$ evaluated in numerical simulation in the split EIA feature.
To guide the eye, a linear fit to the 3rd and subsequent points is provided.
\textbf{(c)}
Pairwise interaction shift $\Delta_I(r)$ for the most shifted eigenstates (green - repulsive and red - attractive). 
Nearest neighbor distance CDF $P_{\text{N.N.}}(r)$ shown in blue on the right y-axis, for the highest Rydberg atom number density attained in the experiment
$n_{\ket{3}} = \SI{9.33e12}{\per\meter\cubed}$.
}
\label{fig:interaction_compare}
\end{figure*}

\subsection{Ionization and Stark effect}
Ionization of the Rydberg atoms produces ions in the cell, which in turn generate a random distribution of electric fields. 
These electric fields cause level energy shifts through the DC-Stark effect.

 \paragraph{Stark effect}
We calculate the DC-Stark effect curve for the $\ket{55P_{3/2}\;m_j=3/2}$ state using the Alkali Rydberg Calculator~\cite{ibali2017}, as presented in Fig.~\ref{fig:interaction_compare}(a) by the orange solid curve. 
It falls in the quadratic regime: $\Delta_S = -\alpha_s E^2/2$ with negative shifts. The polarizability for the state is $\alpha_s = 2\pi\times\SI{0.606}{\alfasiunits}$. We also provide a curve for the $\ket{55P_{3/2}\;m_j=1/2}$ state, with polarizability $2\pi\times \SI{0.718}{\alfasiunits}$ as a reference that it is of similar magnitude. Assuming the quadratic Stark shift formula, we proceed to calculate the electric field needed to cause the measured level energy shifts.
 
\paragraph{Holtsmark}
Ionization of Rydberg atoms creates an ion density $\nion$, which in turn generates a random electric field. 
The field strength follows a Holtsmark distribution~\cite{Pain2020},
which is depicted in Fig.~\ref{fig:interaction_compare}(a) (red, right y-axis).

For the sake of comparison, we take the median of the distribution, $E_{\text{median}} \approx 0.333 \frac{e}{\epsilon_0} \nion^{2/3}$, as the representative electric field causing the measured detuning,
because the position of minima in transmission spectra, like in Fig.~\ref{fig:results}(a), will be determined by the most probable contribution. 
On the contrary, the infinite tail of the Holtsmark distribution will only provide a widely smeared background to observed spectra.

\paragraph{Ion density}
Combining polarizability with the median field, we arrive at the formula connecting $\Delta_\text{comp}$ with $\nion$:
\begin{equation}
\nion = C \Delta_{\text{comp}}^{3/4},\quad\text{where}\quad C \approx 8.75 \left(\frac{\epsilon_0^2}{e^2 \alpha_s}\right)^{3/4}.
\label{eq:ions}
\end{equation}
 Figure~\ref{fig:interaction_compare}(b) presents the ion number density $\nion$, corresponding to the measured
 Rydberg level energy shift $\Delta_\text{comp}$ (c.f. Fig.~\ref{fig:results}(b)) against the Rydberg number density $n_{\ket{3}} = n_{\ket{0}}\bar\rho_{33}$, with $\bar\rho_{33}$ taken from the calibrated simulation in the split EIA feature and $n_{\ket{0}} = \SI{8.46e15}{\per\meter\cubed}$ - from calibrating atomic densities in the split EIA scan.
The results might suggest a threshold process, and to guide the eye, we fitted a line to all but the first 2 points.

The density of ions inferred from the measurements is higher than the Rydberg atoms density.
This could be due to equilibration of collisional and relaxation processes in the vapor, where speeds of ion-generating processes may be larger than speeds of ion-loss processes.

\subsection{Van der Waals interaction}
\paragraph{Interaction curves}
We calculate the energy shift for pair states of the Rydberg state as a function of the interatomic distance $\Delta_I(r)$, for different pairs of angular momentum projection values of the two atoms, $m_{j1}, m_{j2}$, with the Alkali Rydberg Calculator~\cite{ibali2017}, yielding 16 eigenstates.
Depending on the relative orientation of the atomic pair with respect to the axis of the beams, we couple to different eigenstates.
The energy shifts with the extreme positive and negative $\Delta_I$ values are shown in Fig.~\ref{fig:interaction_compare}(c) and denoted by the pair state's total angular momentum projection $M = m_{j1} + m_{j2}$ onto the axis joining the atoms.

\paragraph{N.N. distribution}
The distribution of distance between atom pairs is the nearest-neighbor (N.N.) distance distribution, and for a uniform atom density, this is a Poissonian distribution~\cite{Chandrasekhar1943}. The cumulative distribution function (CDF) $P_{\text{N.N.}}(r)$ of this distribution is visible in Fig.~\ref{fig:interaction_compare}(c) (blue, right y-axis) for $n_{\ket{3}} = \SI{9.33e12}{\per\meter\cubed}$. For this value of $n_{\ket{3}}$, the distribution has a median distance $r_m=\SI{25.74}{\micro\meter}$ corresponding to atoms interacting negligibly, $\Delta_I(r_m) < - 2\pi\times \SI{0.1}{\mega\hertz}$.

\paragraph{Expected value}
Next, we estimate a distance-averaged shift due to van der Waals interactions.
To this end, we calculate the expected value of the most negative interaction curve over the N.N. distance distribution.
For the Rydberg atom number density $n_{\ket{3}} = \SI{9.33e12}{\per\meter\cubed}$ - the highest measured, the
mean energy shift equals $-2\pi \times \SI{0.384}{\mega\hertz}$, which is a value a magnitude smaller than the observations.
This crude calculation overestimates the magnitude of the van der Waals interaction energy shift, as it disregards any averaging over the orientation of the atom, splitting of the $\ket{3}$ state, and modification of the $d_{23}$ dipole moment.

Thus, our experimental results cannot be explained by van der Waals interactions. 
We observe an energy shift of well-defined magnitude, too high to be caused by van der Waals interactions for the estimated atomic densities.

\section{Summary}
\paragraph{What we present, practical applicability}
We developed a general scheme for measuring top-level energy shifts in a four-level ladder setup, based on balancing the two minima of a split EIA in the probe transmission spectrum.
We utilized our method for a robust measurement of interaction-induced level energy shifts. The measurement directly yields the Rydberg level energy shift around the central features.
Thanks to the symmetric nature of the split EIA transmission scan and measuring features close to resonance, the relevant Rydberg atom density is easier to determine.

The ``split EIA balancing'' method is readily deployable in many experimental schemes, which use a four-level ladder setup with a top Rydberg level~\cite{Wade2018, deMelo2016, Carr2013, BLiu2025}.
We also present a framework to explain the observations, which could provide a basis for developing similar methods for more complex level schemes.

\paragraph{What we infer - fundamental and practical impact}
We compared our results with theoretical estimates and found that van der Waals interactions cannot be used to explain our observations, while ionization-induced Stark shifts can. Our results suggest the existence of a threshold below which few ions form, and they influence the system negligibly. Above a threshold, we would observe an amount of ions proportional to the amount of Rydberg atoms, with the ions outnumbering Rydberg atoms.

We challenge the widespread narrative that the interaction-induced level shift is proportional to Rydberg state population, caused by van der Waals interactions~\cite{Weichman2025, XWu2024, Carr2013, Wu2026, Jiao2025, deMelo2016, BLiu2025, BLiu2024, Gambetta2019, He2020, Zhang2024, Ding2024, Marcuzzi2014, Zhang2025, Ma2024}.
Thus, our findings present a step towards modeling bistability as well as theoretically predicting and understanding its occurrence. Employing the correct form of the interactions is especially important in metrology for theoretical assessment of a protocol's performance.

Our method may be meaningful for designing deployable implementations of quantum sensing schemes with Rydberg atoms. It can be used to directly probe the maximum permissible Rydberg density before ionization. Random fields produced by ions are bound to destroy the atomic coherence necessary for these schemes. We believe more work should be dedicated to assessing the performance of proposed sensors in the high Rydberg atom density regime, where these interactions become significant and could limit their capabilities.


\section*{Acknowledgments}
The “Quantum Optical Technologies” (FENG.02.01-IP.05-0017/23) project is carried out within the Measure 2.1 International Research Agendas programme of the Foundation for Polish Science, co-financed by the European Union under the European Funds for Smart Economy 2021-2027 (FENG). This research was funded in whole or in part by the National Science Centre, Poland grant No. 2024/53/B/ST2/04040.

\section*{Author Contributions}
T.P. and B.K. contributed equally to this work.






\section*{Data availability} Data underlying the results presented in this paper are available in Ref.~\cite{data}.

\bibliographystyle{apsrev4-2}
\bibliography{refs}
\end{document}